# Adaptive Multimodal Music Learning via Interactive-haptic Instrument

Yian Zhang    Yinmiao Li    Daniel Chin    Gus G. Xia

Music X Lab
New York University, Shanghai
{ yian.zhang, yl4121, nq285, gxia } @nyu.edu

## ABSTRACT
Haptic interfaces have untapped the sense of touch to assist multimodal music learning. We have recently seen various improvements of interface design on tactile feedback and force guidance aiming to make instrument learning more effective. However, most interfaces are still quite static; they cannot yet sense the learning progress and adjust the tutoring strategy accordingly. To solve this problem, we contribute an adaptive haptic interface based on the latest design of haptic flute. We first adopted a clutch mechanism to enable the interface to turn on and off the haptic control flexibly in real time. The interactive tutor is then able to follow human performances and apply the "teacher force" only when the software instructs so. Finally, we incorporated the adaptive interface with a step-by-step dynamic learning strategy. Experimental results showed that dynamic learning dramatically outperforms static learning, which boosts the learning rate by 45.3% and shrinks the forgetting chance by 86%.

## 1. INTRODUCTION
Learning to play an instrument is intrinsically multimodal. It usually involves learning music notations via the visual system, memorizing the tones via the auditory system, and mastering the performance skills via the motor system. Though visual and auditory interfaces, such as sheet music and recordings, have long been used to assist music learning, haptic interfaces that are able to reproduce motion expressions have just been invented in the recent years. [2, 3, 4, 5, 8]. In general, haptic interfaces offer guidance via tactile or kinesthetic perception: tactile perception is vibrations or pressure conveyed through the skin, while kinesthetic perception is receptors in muscles and tendons that allow us to feel the pose of our body [6].

Haptic interface is a revolutionary technology for instrument learning, as people no longer have to *infer* the performance motions from score or sound, but rather can learn the motions *explicitly* and *directly*. From a cognitive perspective, explicitly learning performance motions can reduce the errors brought by motion inference and hence form a firmer memory. In addition, only with the correct motion sequence can one directly play a piece of music via an instrument, while the score and sound are high-level abstractions that can only help indirectly. Therefore, learning performance motions directly can help people master a piece of music more quickly, especially for beginners who have no prior knowledge of music.

In practice, we indeed have seen several promising prototype systems proving that haptic guidance improves musical motor learning, especially when haptic guidance is combined with acoustic and visual guidance. Grindlay [3] applied haptic guidance to learn drum kick sequences, Huang et al. [4] developed a wearable tactile glove to learn the piano, and Fujii et al. [2] developed a haptic device to learn Theremin. In a recent study, Xia et al. [8] successfully applied haptic-guided flute tutoring to help beginners learn short folk songs and significantly improved the learning rate compared to learning via videos.

However, as far as we know, the interfaces are still quite *static*. They always use the same strategy to enforce human motions repeatedly, since both hardware and software design lacks the flexibility and intelligence to adjust the guidance according to the student's learning progress. As discovered in [8], many learners wish the haptic interface provide some freedom for trial-and-error, especially towards the end of the learning when they feel most motions have been mastered. If the weakness of traditional instrument learning is bringing in too much uncertainty, the weakness of current haptic learning is offering too little freedom. From a machine learning perspective, this difference is analogous to the one between unsupervised learning and fully supervised learning. We aim to find a balanced zone of controlled freedom (as shown in Figure 1), where learners can actively test their skills under the guidance, rather than being guided passively again and again.

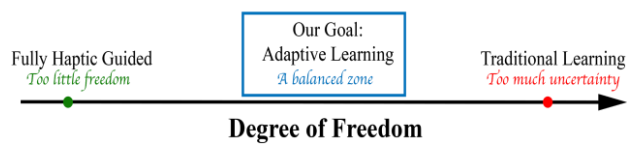

**Figure 1. A big picture of haptic guidance.**

In this paper, we developed an adaptive haptic interface based on the latest design of haptic flute [8]. To be specific, we contributed 1) a hardware design with a clutch mechanism to enable the finger rings to turn on and off the haptic control flexibly in real time, 2) an interactive algorithm to follow human performances and apply the "teacher force" only when the software instructs so, and 3) a step-by-step passive-to-active dynamic tutoring strategy that contains three modes: the mandatory mode, the hinted mode, and the adaptive mode. The mandatory mode strictly controls the finger positions, the hinted mode applies force on the note onsets but does not enforce finger positions, and the adaptive mode only reacts to the player's mistakes. This dynamic tutoring strategy is inspired by the Scaffolding method [7] in Education, which interactively assists the learning while gradually removing the guidance as learners get more proficient. Experimental results showed that our dynamic learning strategy dramatically outperforms static learning, which boosts the learning rate by 45.3% and shrinks the forgetting chance by 86%.

## 2. METHODOLOGY
We present the hardware design of our haptic interface in Section 2.1, describe the adaptive algorithm and the three learning modes in Section 2.2, and finally introduce the dynamic tutoring strategy in Section 2.3.

## 2.1 Hardware Design

We first introduce the baseline hardware design in Section 2.1.1 and then present our improvements from Section 2.1.2 to Section 2.1.4.

### 2.1.1 Point-to-point control

The point-to-point control is the baseline design developed by Xia et al [8]. The design is shown in Figure 2, in which six servos (rotation arms connected to finger pegs) are used to guide the positions of fingers. Fingers, finger pegs, and rotation arms of servos are "locked" on one another all the time, so that a one-to-one correspondence is created between: (A) servo arm angles, and (B) finger positions. Although this design can provide firm haptic guidance, fingers are strictly constrained by the device and hence have no freedom of movement.

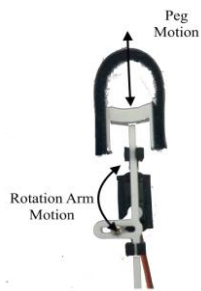

**Figure 2. The baseline design.**

### 2.1.2 Point-to-range control

The new design uses linear servo actuators to replace traditional servos and maps every position of the servo arm to a *range* of finger position rather than a fixed point. As shown in Figure 3, the servo arm is no longer locked on the finger peg but put into a sliding rail so that the finger (with the finger peg) can move up and down freely within a dynamic range. This range = ½ servo arm track - ½ servo arm width, and the center point of the range is decided by the height of the servo arm. In this way, the point-to-range control offers the fingers some freedom, which is a key requirement of adaptive learning.

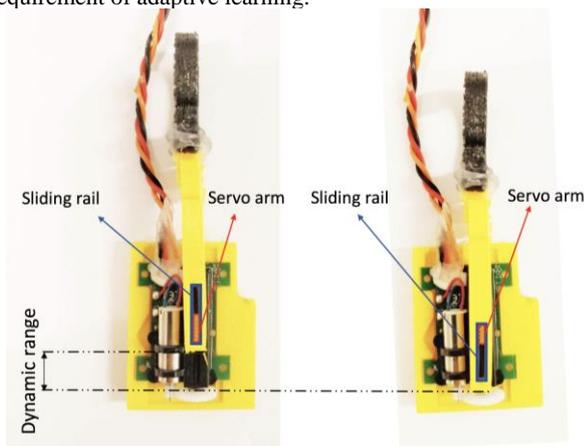

**Figure 3. An illustration of point-to-range design. (Detached state. Notice that the servo arm did not move)**

### 2.1.3 Clutch mechanism

Based on the point-to-range control system, we design a "clutch mechanism", which enables the device to freely switch between two states: the *detached* state and the *attached* state. When the device is in the detached state, the servo arm stays in the middle of its track, and fingers can move freely within the dynamic range as shown in Figure 3. To switch to the attached state, the servo arm moves to the top of the sliding rail to fix a finger at the highest point or to the bottom of the sliding rail to fix a finger at the finger hole. Figure 4 shows the clutch mechanism.

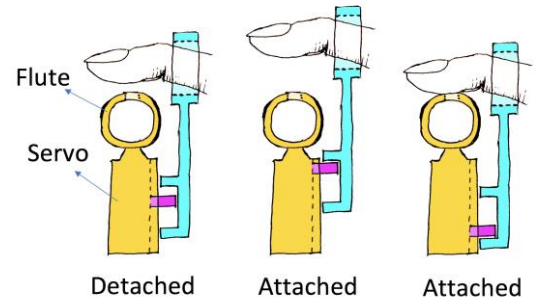

**Figure 4. An illustration of the clutch mechanism.**

### 2.1.4 Other system components

Three other important hardware components are the servo shells, zipper-style flexible finger pegs, and the adjustable paddings. Figure 5 shows the overall design of the haptic interface.

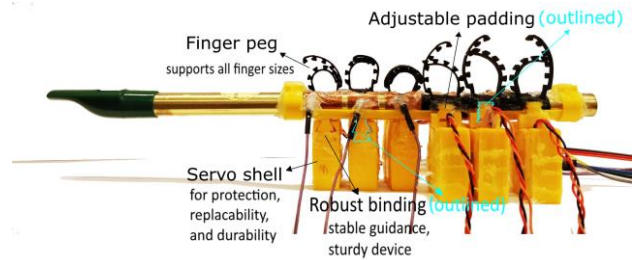

**Figure 5. The overall design of flute device.**

**Servo shell**: every servo is integrated into a box-shaped shell. The shell protects the servo from direct impacts and environmental hazards, only leaving a small window for the sliding rail and the cables. We design a robust binding between the flute and the shell for stable haptic guidance and a durable device.

**Zipper-style flexible finger peg**: 3d-printed with TPU 95A, the two bendable racks can interlock at different lengths, accommodating to the learner's finger size.

**Adjustable padding**: easily configures the height offset of each finger peg.

## 2.2 Learning Modes and Algorithms

We developed three multimodal learning modes. Given a piece of music to learn, all three modes play back the music at a constant tempo and provide synchronized haptic guidance. The difference among the three modes lies in how the haptic guidance is triggered.

### 2.2.1 Mandatory mode

This is the mode developed and tested in the baseline design [8]. In this mode, positions of the fingers are completely controlled by the device. We achieve the same effect using the new device, by always using the attached state.

### 2.2.2 Hinted mode

The hinted mode exerts a moderate force on the finger for every note onset but does not enforce finger positions. This new function is made possible by the clutch mechanism: the finger pegs return to the detached state immediately after the note onsets. In addition, the duration of the force is set to be very short in this mode so that it barely moves the finger but only provides a "hint". Therefore, learners can no longer 100% rely on the device passively, but must actively move their fingers. In other words, the hinted mode offers fingers some freedom for trial-and-error.

*2.2.3 Adaptive mode*

The adaptive mode enables a more active learning by only taking actions when learners make mistakes. We use capacitive sensors to detect finger motions and compare the learner's performance with the ground truth (the predefined score) in real time. For a note whose score time is *t*, the algorithm would report a mistake if no correct pitch is observed from $t$ to $t + \Delta t$. In practice, we set $\Delta t$ to be 200ms.

## 2.3 A Dynamic Learning Strategy

A passive-to-active dynamic learning strategy is designed by combining the three learning modes introduced in the last section: learners first use the mandatory mode to lay the groundwork and gradually move to the hinted mode and the adaptive mode as they become more proficient. This dynamic learning strategy is inspired by the theory of *instructional scaffolding* and the *zone of proximal development* (ZPD) [7] in Education Science. This theory classifies all the skills into three zones, as shown in Figure 6, where ZPD refers to the skills that are too difficult to be learned independently but can be learned under guidance.

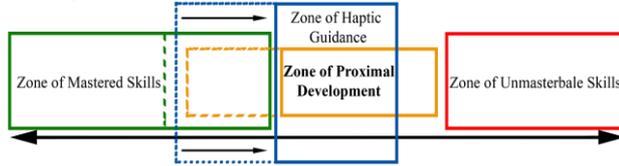

**Figure 6. Dynamic haptic learning strategy in the framework of instructional scaffolding.**

Previous studies on effective haptic instrument learning [8] have proved that the zone of haptic guidance (the blue area) overlaps with the ZPD. However, so far as we know, existing haptic learning methods are all static, which means the haptic zone does not change over the learning procedure when the ZPD moves. Our dynamic learning strategy can be understood as interactively pushing the left boundary of the haptic guidance zone to the right as more skills are mastered. Our future studies will focus on how to push the right boundary to master more advanced performance skills.

## 3. EXPERIMENTS

To validate the effectiveness of the proposed dynamic learning strategy, we conducted a quantitative study to compare it with our baseline algorithm, the static learning strategy. Using the same device, we reproduced static learning by solely using the Mandatory Mode.

### 3.1 Music to Learn

We re-used the two pieces composed in [8] for our study. The two songs are designed to be of similar difficulty with the same pitch range, the same number of intervals, and the same amount of finger movements. Figure 7 shows their scores.

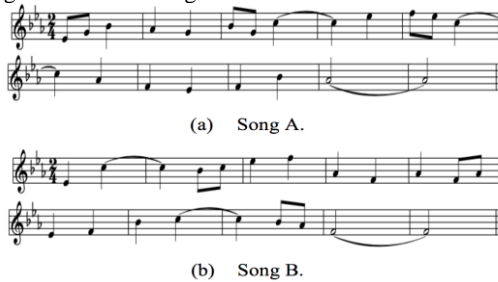

**Figure 7. The two songs to be learned.**

### 3.2 Participants

Eighteen paid participants (7 males and 11 females) between the age of 19 and 27 participated in the study. All participants had no experience playing the flute and reported no familiarity with the composed pieces. Also, no participants in this experiment were involved in the experiment of [8], so they were all unfamiliar with haptic devices. All but two of the eighteen participants completed both songs.

### 3.3 Design

The experiment employed a 2×2 within-subject factorial design. The independent variables were the learning method (dynamic learning, static learning) and learning pieces (Song A, Song B).

Each participant played both song A and song B: one learned through dynamic learning, and the other learned through static learning. We cycled through all four permutations (of song-choice and song-learning method combination) four times to produce our 16 data points.

**Table 1. Counterbalancing independent variables.**

| Subjects | First trial | Second trial |
|---|---|---|
| 25% | Song A with dynamic | Song B with static |
| 25% | Song A with static | Song B with dynamic |
| 25% | Song B with dynamic | Song A with static |
| 25% | Song B with static | Song A with dynamic |

### 3.4 Task and Procedure

The procedure consisted of 4 steps: the pre-training step, the learning & testing step, the interview & forgetting step, and the re-exam step. The participants were allowed to give up at any time in any step.

**Pre-training**: In this step, we taught participants to play a basic scale on the flute through the range of the song they would be playing. This step ensures they had the baseline ability to play.

**Learning & testing**: At this step, participants learn the first piece using the first learning method (dynamic/static) with audio playback. Participants can test their skills at any time when they feel confident enough. In case a test is failed, they are free to shift back and learn again and again. A test is passed if the participant can reproduce the piece with the correct pitch sequence via a normal flute (not equipped with any haptic device).

**30-mins Interviewing & forgetting**. After the participants mastered the first song, we interviewed them with several questions and then played tongue twisters with them to accelerate the forgetting process on their muscle and musical memory. The interview questions and the games to play were the same for all participants. The duration of this step was fixed to be 30 minutes.

**Re-exam**: The participants were asked to reproduce the learned piece again without haptic guidance. One is regarded as *forgetting* the piece if he or she failed in the re-exam. This re-exam procedure is to test whether the piece is memorized firmly. After the four steps, we repeated the whole process with the second learning method (static/dynamic) to learn the second piece.

### 3.5 Results and Discussion

*3.5.1 Learning efficiency*

Figure 8 summarizes the comparison of learning efficiency between the two learning methods, where we see the dynamic learning strategy outperforms the baseline significantly. Here, the x-axis represents the time spent on dynamic learning and the y-axis represents the time spent on static learning. We see that all but three points are above the y = x line, which shows that most participants learned the piece faster using dynamic learning. Excluding the people that failed to learn the pieces, the proposed dynamic learning method achieved significant improvement, with $p < 0.005$ by pairwise t-test and an average of 45.3% increment in the learning rate (in terms of percentage of a piece per minute).

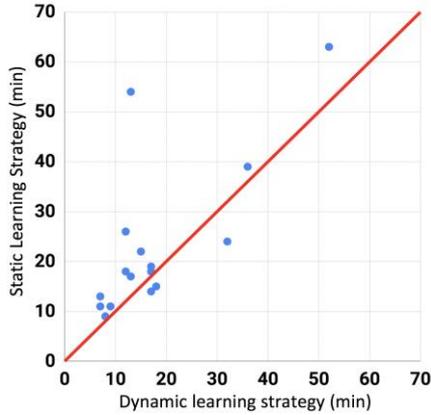

**Figure 8. A comparison of learning time between our method and the baseline method (above the red line is where dynamic learning is faster).**

### 3.5.2 Forgetting chance

Table 2 shows the comparison of the forgetting chance between the two learning strategies. "Forgetting chance" is defined as the percentage of participants who failed to reproduce the piece correctly after 30 minutes. 7 out of 16 participants using static learning forgot the piece. In contrast, only 1 participant who used dynamic learning forgot the piece. In other words, dynamic learning leads to a firmer memory and proportionally shrinks the forgetting chance by 86% (from 43.75% to 6.25%).

**Table 2. A comparison of forgetting chance.**

| Learning Strategy | Number of Learning Success | Number of Forgetting after 30 minutes | Forgetting Chance |
|---|---|---|---|
| Static | 16 | 7 | 43.75% |
| Dynamic | 16 | 1 | 6.25% |

### 3.5.3 Long-term tracking results

We also conducted a long-term experiment aiming to compare the forgetting rates of the two learning strategies over a longer period of time, five days. This experiment concentrated on one person, using the static learning strategy for the song A and dynamic learning strategy for song B. After the learner mastered the songs, we examined the forgetting rate every day afterward. If the exam result suggested the user forget the piece, even partially, we let him learn it again each time.

The long-term forgetting ratio results are shown in Figure 9, from which we see that dynamic learning enhances long-term memory compared to static learning. Here, the x-axis represents the time lag in days and the y-axis represents the forgetting ratio (= the number of forgotten notes / the number of total notes) (so smaller values mean better results, and 0 value means 100% memorized) The solid blue line represents the forgetting curve of static learning, and the dotted red curve represents the forgetting curve of dynamic learning. In day 1, the user learned the song A in 15 mins and song B in 9 mins. In every following day, the forgetting rate of song A is higher than that of song B, as can be seen in Figure 9. Also, it always took the user less time to pick up song B.

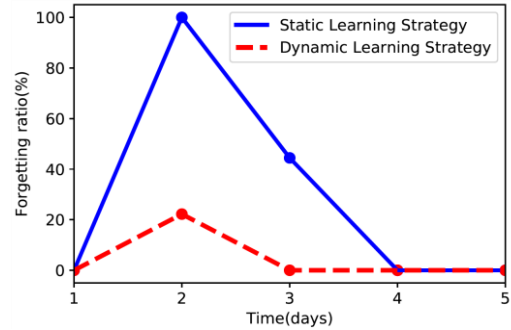

**Figure 9. A comparison of long-term memorization (smaller means better).**

### 3.5.4 Interview and discussion

Here, we report several interview questions and some representative answers to help gain a deeper understanding of the adaptive interface.

**Q1: How do you feel about learning a piece using the static learning strategy compared to using the dynamic learning strategy? Would you prefer learning with haptic guidance or traditional learning by looking at the score?**

- *"Dynamic learning is more interesting because of the real-time feedback. I feel more motivated to learn, and to break through. When I used the static mode, I was nervous. In comparison, It felt really good when I hit notes correctly in the adaptive mode." ……*

- *"I prefer the mandatory mode because its behavior is predictable, and I would know when there will be a note. The adaptive mode is too uncertain. I got different feedback for every round of practice, and I got confused."*

- *"I prefer the dynamic strategy. The static mode lacks a step-by-step process. I can't try on my own and I don't know if I already learned the piece correctly in the static mode. Also, I would like to learn with breathing and hand position learning separated."*

- *"I'd prefer haptic guidance over learning via score. Haptic guidance is more motivating, and it provides a sense of safety because it is like someone teaching you."*

- *"I prefer learning with the haptic guidance. It will be even better if I am allowed to look at the score when learning by the haptic device."*

**Q2: Any comments on the three learning modes or suggestions for improvements?**

- *"The hinted mode and the adaptive mode are more comfortable for me. They help me feel the change in the music piece. It feels like a flute master is guiding me, reminding me of details. On the other side, I would suggest increasing fault tolerance rate, so I wouldn't feel disrupted when there are many errors."*

- *"Besides the finger part, I need tutoring on breathing, too. Also, maybe you can remove the hinted mode because its function overlaps with the rest two modes too much. Maybe you can add visual guidance to remind user of tempo and breathing changes.*

*You can also add more expression tutoring, so there are emotions in the performance."*

**Q3: Please use one to three words to describe your feeling about the overall learning experience**

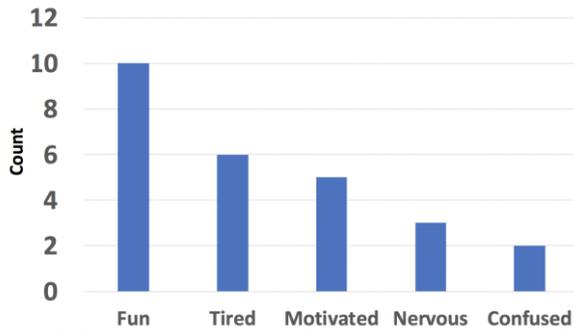

**Figure 10. The top five words used to describe the overall learning experience.**

### 3.5.5 More discussions

During the experiment, we notice another interesting phenomenon worth discussion: *Previous music experience strongly affects haptic learning*. Those with music learning experience showed either the best or the worst performance. Some of them picked up haptic learning extremely fast, while some others struggle to make progress. Our explanation for the latter is that they tended to rely more on their musical knowledge and experience than on the haptic guidance. One of these participants struggled the learning by trying to decode and memorize the mapping between sound to finger position, but learned the piece much faster after she decided to let go and feel the muscle memory.

## 4. TOWARDS MORE ADAPTATION

In this Section, we present two more hardware designs, which generalize the idea of the clutch mechanism to achieve more adaptation. We have already seen that one key feature of adaptive learning is to constrain less while learners are making progress. Therefore, more adaptation means more degree of freedom to the fingers during the haptic guidance, which has the potential to help users learn more advanced performance techniques and even help users generalize the techniques across different instruments. In Section 4.1, we present the C-ring design, and in Section 4.2, we show the Magic Gloves.

### 4.1 The C-ring Design

Figure 11 shows the C-ring design. To perform guidance, servo motors drive the rotary movement of the "C" shape levers to lift up or push down the corresponding fingers. Such design sets the fingers free from the finger pegs and finger straps, allowing fingers to move both vertically and horizontally. Actually, the rings only need to touch the finger when they apply instructions. Under the mandatory mode, the rings rotate, move the fingers, and hold still. Under the hinted and adaptive mode, the rings rotate, touch the fingers, and immediately rotate back to the mutual position.

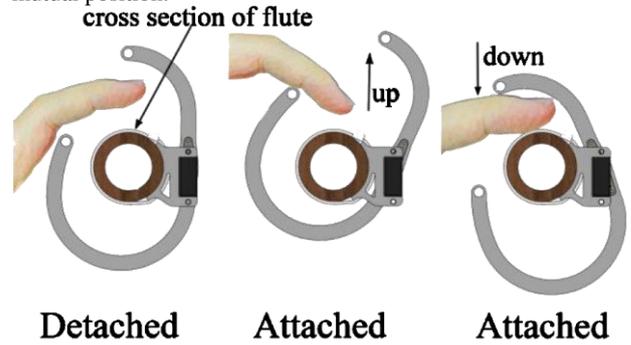

**Figure 11. An illustration of a single lever**

### 4.1.1 Hardware and electronic

Figure 12 shows a global view of the entire device. The device contains six 3D printed servo holders that bind to the flute with Nylon cable ties. Six 3D printed "C"-shape lifting levers, used for directing users' fingers, enclose the six holes on the flute by attaching to the servos (as shown in Figure 11). The motors are connected to Arduino Pro Mini for signals. The connecting wires from the Arduino board to the computer and the external power supply harm the wielding experience, so we make the device wireless. We adopt Bluetooth Module (HC-05) to receive signals from any computers connected to the module and use battery connected to a boost converter module for power supply. In addition, six pieces of foil are attached to the finger holes for capacitive sensing in order to realize the adaptive mode.

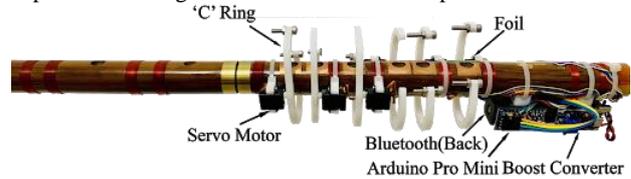

**Figure 12. An illustration of the entire haptic interface with the C-ring design.**

### 4.1.2 Merits and demerits

The rotary movement makes full use of the torque provided by the servo motor, which leads to more powerful guidance. Servo motors and Arduino Pro Mini are inexpensive and thus suitable for prototyping. The design brings more degree of freedom, but user testing suggests it costs more time for learners to cooperate with the guidance.

### 4.2 The Magic Gloves

The design of the Magic Gloves generalizes the concept of adaptive learning to other instruments. This is achieved by relocating the interface from the flute to the hands. The idea of wearable haptic device is not new. (For example, a tech company [1] developed CyberGrasp back in 2016.) Our Magic Gloves are tailored for instrument learning and present an inexpensive solution with minimal mechanical complexity. Figure 13 shows our prototype design.

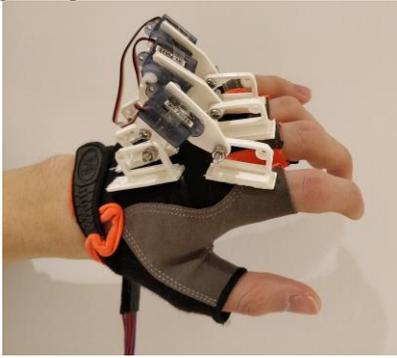

**Figure 13. An illustration of the Magic Gloves**

### 4.2.1 Benefits of wearable interface
Many instruments share similar performance techniques, but most haptic interfaces are instrument-specific. With a wearable interface, our current software to control the fingers will be highly reusable. As our first prototype, the current Magic Gloves only provide vertical haptic guidance for each finger. This is fine for the flute since the flute has a *static* finger-to-hole mapping. That is often not the case for most other instruments (e.g., saxophone and piano), so further work is needed.

### 4.2.2 Haptic learning method and experience
Because the clutch point-to-range mechanism is intrinsically supported by the stretchability of the glove material, we apply the same control algorithm that we use for the linear servo device. Therefore, the haptic learning procedure is exactly the same between the linear servo device and the Magic Gloves.

Regarding the learning experience of the prototype, we find that the Magic Gloves provide less clarity than our flute-based device does. That is likely due to the "indirectness" of the haptic guidance. (The flute-based device controls the parts of fingers right next to the finger holes, while the Magic Gloves control the third knuckles. See Subsubsection 4.2.3.)

### 4.2.3 Mechanism

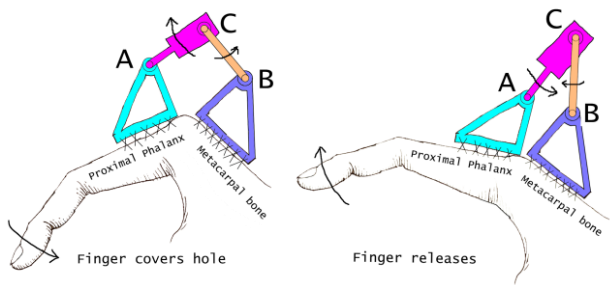

**Figure 14. Magic Gloves controlling the index finger (glove omitted)**

As is seen in Figure 14, the hardware is composed of four moving parts (cyan, magenta, orange, blue) connected by A, B, and C. A and B are simple bearings; C is the torque output of the servo (magenta + orange).

For convenience, let us name the distance between A and B as "$AB$"; the angle formed by AC and BC as "$\angle ACB$"; the angle formed by the proximal phalanx and the metacarpal bone as "$\angle knuckle$". When $\angle ACB$ increases, $AB$ will increase, and $\angle knuckle$ will decrease, and the finger will be pushed down. In this way, we can let the servos control the finger motions.

## 5. CONCLUSION AND FUTURE WORK
In conclusion, we developed an adaptive interactive-haptic flute system as a novel method for haptic-supported learning of instruments. This system tackles the problem of being too static by introducing the clutch mechanism. Utilizing the new hardware device, we developed both the hinted mode and the adaptive mode. Furthermore, we implemented a dynamic learning strategy for our system. Our experiment shows that the learning rate from the group using dynamic learning strategy is 45.3% higher than the group using the static learning strategy (the baseline). Moreover, the forgetting chance from the group using dynamic learning strategy is 86% lower than the group using the static learning strategy. Above all, we consider this study an interesting path towards the integration of both human-human and human-computational learning of musical instruments through haptic technologies and a major step towards haptic-based music pedagogy and performance assessment.

Though the latest system has made promising progress on flute tutoring, there are still limitations remaining and further improvements to make. Firstly, the adaptive mode is tempo fixed, which means that users have to follow by the exact same tempo as the given reference. We will continue to develop an algorithm that tracks user playing regardless of the original tempo and provides feedback based on smarter error detection. Secondly, visual guidance proves to help with multimodal music learning. Hence, we will develop the system with both visual and haptic guidance. Last but not least, we want our device not only to be useful for beginners but also help intermediate learners learn more advanced techniques.

## 6. ACKNOWLEDGMENT
We want to thank all the participants in our study. We would also like to thank XingDong Yang, Rodolfo Cossovich, Wen Yin, Minchuan Zhou, Rundong Jiang, and Ruitang Chen for their timely feedback and recommendations.